\documentstyle[12pt]{article}
\begin{document}

\title{ Expansion in the Width: the Case of Vortices
\thanks{Paper supported
in part by the grant KBN 2 P302 049 05.}}

\author{by\\
\\
H. Arod\'z   \\
\\
Institute of Physics, Jagellonian University,
Cracow \thanks{Address: Reymonta 4, 30-059 Cracow, Poland.}
        \thanks{E-mail: ufarodz@ztc386a.if.uj.edu.pl} }

\date{  $\;\;$}
\maketitle

\thispagestyle{empty}
 $\;\;\;$ \\
{\bf Abstract} \\
We construct an approximate solution of field equations in the Abelian Higgs
model which describes motion of a curved vortex. The solution is found to
the first order in the inverse mass of the Higgs field with the help of
the Hilbert-Chapman-Enskog method. Consistency conditions for the
approximate solution are obtained with the help of a classical Ward
identity.  We find that the Higgs field of the curved vortex of the
topological charge $n \geq 2$ in general does not have single n-th order
zero. There are two zeros: one is of the (n-1)-th order and it follows a
Nambu-Goto type trajectory, the other one is of the first order and its
trajectory in general is not of the Nambu-Goto type. For $|n|=1$ the
single zero in general does not lie on Nambu-Goto type trajectory.

$\;\;$ \\
February 1995     \nopagebreak    \\
TPJU-3/95

\pagebreak

\setcounter{page}{1}
\section{Introduction}

In the recent paper \cite{1} we have shown how to construct approximate
domain wall solution with the help of Hilbert-Chapman-Enskog method
\cite{2} for solving equations with singular perturbations. The solution
has the form of a perturbative series in positive powers of the width
of the domain wall. Crucial role is played by certain consistency conditions
for the perturbative expansion. Their appearence is related to existence
of zero modes.  In order to satisfy the consistency conditions we
have had to allow for the possibility that the so called co-moving
coordinate system is tied to an auxilliary membrane instead to zeros
of the scalar field. Our approach differs essentially from earlier attempts
to use the expansion in the width for domain walls.

In the present paper we would like to apply this same method to obtain
a solution in the Abelian Higgs model describing a generic
non-selfinteracting vortex.
By the non-selfinteracting vortex we mean a vortex such that no its parts
which are distant along the vortex are spatially close to each other.
This excludes, e.g., spikes and selfintersections. It turns out that the
method works quite well also in the case of vortices. The calculations
we report in the present paper are parallel to the ones in \cite{1}.
Therefore, we do not repeat here general motivation and comments
about the course of calculations. In \cite{1} one can also find a more
extensive list of references including also papers on dynamics of
vortices.

The vortex in the Abelian Higgs model is the classic example of a vortex.
It is called  local because its fields are different from
their vacuum values only in a close vicinity of a single line -- far from
this line the fields are exponentially close to the vacuum values.
Thus, there are no long range forces acting on pieces of the vortex.
For this reason the dynamics of such a vortex is probably simpler than
dynamics of a global vortex with its long range Goldstone field. This is
the reason we have chosen to apply the new method first to the local
vortex. Unfortunately, the Abelian Higgs model is not simple on the
mathematical side. It involves six real fields ($Re\Phi, Im\Phi, A_{\mu}$)
which obey a non-linear set of equations. Moreover, except for the special
case of Bogomol'nyi limit there are present two length scales, given
by the inverses of masses of the scalar and vector fields. Even in the
simplest case of a straight-linear, static vortex the exact analytical
form of the corresponding solution is not known. Therefore, one should be
prepared for rather cumbersome calculations.

The plan of our paper is the following. In the next Section we introduce the
co-moving coordinate system and we write the field equations in the
new coordinates. Most of this material is not new. We quote it here for
convenience of the reader and in order to fix our notation. Moreover, we
would like to emphasize the fact that the co-moving coordinate system
apriori is not tied to position of zeros of the Higgs field. The
approximate position of the zeros are determined at the end, from the
constructed approximate solution of the field equations. In Section 3
we find the zeroth order term in the expansion in the width of the vortex.
By the width we mean the inverse mass of the Higgs field.
We also check the consistency conditions appropriate for the zeroth
order and characteristic for the Hilbert-Chapman-Enskog method.
Section 4 is devoted to the first order solution and the relevant
consistency conditions. Section 5 contains final remarks.
In Appendix  A we derive a classical
Ward identity which helps to calculate troublesome integrals encountered
in Section 4. In Appendix B we give a list of constants appearing in
Section 4.

\section{Equations for the scalar and vector fields in the co-moving
 coordinates}

We take the Lagrangian of the Abelian Higgs model in the following form
\begin{equation}
L = D^{\mu}\Phi^{\ast} D_{\mu}\Phi - \frac{1}{4} F^{\mu\nu}F_{\mu\nu}
- \frac{\lambda}{4} ( |\Phi|^{2} - \frac{ M^2}{\lambda} )^2,
\end{equation}
where
\[  D_{\mu}\Phi = \partial_{\mu} \Phi + i q A_{\mu} \Phi, \;\;
  D_{\mu}\Phi_{\ast} =  ( D_{\mu}\Phi)^{\ast} =
  (D_{\mu})^{\ast}(\Phi)^{\ast},   \]
The Minkowski space-time metric is chosen as
$(\eta_{\mu\nu}) = diag(1,-1,-1,-1)$. The corresponding field equations are
\begin{equation}
 D^{\mu} D_{\mu}\Phi + \frac{\lambda}{2}\Phi\;
  ( |\Phi|^{2} - \frac{M^2}{\lambda} ) = 0,
\end{equation}
\begin{equation}
\partial^{\mu}F_{\mu\nu} = i q ( \Phi^{\ast} D_{\nu}\Phi -
 \Phi  D_{\nu}\Phi^{\ast}).
\end{equation}

The vacuum fields are characterised by the following conditions
\begin{equation}
|\Phi| = \Phi_{0} \equiv \frac{M}{\sqrt{\lambda}}, \;\; D_{\mu}\Phi =0.
\end{equation}
The masses of the Higgs  and the vector particle are equal to
$ m^{2}_{H}=  M^2, \; m^{2}_{A} = \frac{2 q^2}{\lambda} m^{2}_{H}$,
respectively. We consider the case $ \frac{2 q^2}{\lambda} \leq 1$  which
corresponds to  superconductors of the second kind.

In the case of single vortex considered in this paper, the fields
 $\Phi, \; A_{\mu}$ at each instant of time have the vacuum values in
the whole space $R^{3}$ except for a vicinity of a line $C$ where
the energy density is significantly different from zero. We do not
assume that the Higgs field vanishes precisely on this line.
The  world-sheet of the line $C$ in Minkowski space-time we shall denote
by $\Sigma$. The line $C$ we will call the string, in anticipation of
the result of the next Section, where we find that in the framework of
$1/M$ expansion $C$ has to obey an equation identical
in its form with Nambu-Goto equation for a relativistic string. The string
is used to define the coordinate system co-moving with the vortex.
To this end, we parametrize  $\Sigma$ with two parameters
$\tau,\sigma$
\[ [Y^{\mu}(\tau,\sigma)] \in \Sigma.  \]
The parameter $\tau$ is by assumption time-like, i.e.
$Y^{\mu}_{,\tau}  Y_{\mu,\tau} > 0$, while  $\sigma$ is the space-like one,
i.e. $Y^{\mu}_{,\sigma}  Y_{\mu,\sigma} <0$. In the following we use a more
compact notation: $(\tau,\sigma) = (u^a),\; a=0,3,$ with $a=0\; (a=3)$
corresponding to $\tau \;(\sigma)$. The co-moving coordinates are defined
in a vicinity of the string world-sheet  $\Sigma$ by the following
formula
\begin{equation}
x^{\mu} = Y^{\mu}(u^a) + \xi^i\; n^{\mu}_{i}(u^a),
\end{equation}
where $x^{\mu}$ are Cartesian, laboratory frame coordinates in Minkowski
space-time; $i=1,2$;$\;  n_{i}(u^a)$ are two orthonormal space-like
four-vectors, orthogonal to  $\Sigma$ at the point $[Y^{\mu}(u^a)]$
in the covariant sense, i.e.
\begin{equation}
Y^{\mu}_{,a}(u) \; n_{i\mu}(u) =0, \;\; n_{i}^{\mu}(u) n_{k\mu}(u) =
 - \delta_{ik}.
\end{equation}
The conditions (6) do not fix the four-vectors $n_i$ uniquely -- there is
still a freedom of local (i.e. $(u^a)$-dependent) SO(2) rotations acting on
the indices $i,k$. The coordinates $\xi^1,\xi^2$ vanish on the string. We
shall denote the full set of co-moving coordinates by ($\zeta^{\alpha}$):
$\;(\zeta^{\alpha}) = (u^a, \xi^i), \;$  $\alpha$=0,3,1,2. The co-moving
coordinate system has been used in numerous papers, see, e.g.,
\cite{3,4,5,6}, with the difference that in those papers it was attached
to a line on which the Higgs field was assumed to vanish.

The next step is to write Eqs.(2),(3) in the co-moving coordinates. It is
convenient to introduce several mathematical objects:  \\
- the metric $g_{ab}$ on $\Sigma$,
 \[g_{ab}=   Y^{\mu}_{,a}  Y_{\mu,b};\]
- the extrinsic curvature coefficients,
\[ K^{i}_{ab} =  - n^{\mu}_{i}Y_{\mu,ab}; \]
- the torsion coefficients
 \[ \omega_a = \frac{1}{2} \epsilon^{ik} n_{i} n_{k,a}, \]
where $\epsilon^{ik}$ is the 2-dimensional totally antisymmetric symbol
($\epsilon^{12} =1$). The torsion and extrinsic curvature coefficients
depend on the choice of the four-vectors $n_{i}$. In general it is not
possible to adjust them in such a manner that the torsion vanishes. This
can be seen from Ricci equation
\[ \partial_{a}\omega_{b} -  \partial_{b}\omega_{a} = \epsilon^{ij}
K^{ic}_{a} K^{j}_{bc}, \]
where  $ K^{ic}_{a}\equiv K^{i}_{ab} g^{bc}$. Vanishing of the torsion
would imply that the extrinsic curvatures at a given point could not be
arbitrary.

The Minkowski space-time metric tensor in the new coordinates has the
following form
\begin{equation}
[G_{\alpha\beta}] = \left[ \begin{array}{cc}
G_{ab} & G_{ak} \\
G_{ia} & G_{ik}
\end{array}
\right],
\end{equation}
where
\[ G_{ab} = g_{ab} + 2 K^{i}_{ab} \xi^{i} + K^{ic}_{a} K^{j}_{bc}
 \xi^{i}\xi^{j} - \xi^{i}\xi^{i} \omega_{a}\omega_{b}, \]
\[ G_{ai} = G_{ia} = \epsilon^{ik} \xi^{k} \omega_{a}, \]
\[ G_{ik} = - \delta_{ik}. \]
It follows from these formulae that
\[ \sqrt{-G} = \sqrt{-g}\; h(u^a, \xi^i), \]
where $G=det[G_{\alpha\beta}],\; g=det[g_{ab}]$, and
\begin{equation}
h(u^a,\xi^i) = 1 + K^{ia}_{a}\xi^{i} +
\frac{1}{2}(K^{ia}_{a} K^{jb}_{b} - K^{ib}_{a} K^{ja}_{b}) \xi^{i}\xi^{j}.
\end{equation}
We shall also need components of the inverse metric tensor
 $[G^{\alpha \beta}]$. They are given by the following formulae:
\begin{equation}
G^{ab} = h^{-2}[ g^{ab}(1+\xi^{i}K^{ic}_{c})^2 - 2 \xi^{i} K^{iab}
(1+\xi^k K^{kc}_{c}) + \xi^{i} \xi^{j} K^{ia}_{c} K^{jbc}],
\end{equation}
\begin{equation}
G^{ai} = \epsilon^{ik} \xi^{k} \omega_{b} G^{ab},
\end{equation}
\begin{equation}
G^{ik} = - \delta^{ik} + ( \delta^{ik}  \xi^{l}\xi^{l}
-    \xi^{i}\xi^{k}) \omega_{a}\omega_{b} G^{ab}.
\end{equation}
We shall write the Higgs and Maxwell equations (2),(3) in the co-moving
coordinates using rescaled variables:
\[ \xi^{i} = \frac{\sqrt{2}}{M} s^{i}, \;\; \Phi(u^a, \xi^i) = \Phi_{0}
\phi(u^a,s^i). \]
Also the vector field $A_{\mu}$ is transformed to the new coordinates. It
has the following components
\[ (A_{\alpha}) = (A_{a}, A_{i}),   \]
where $ a=0,3,\;\; i=1,2,$ and
\[ A_{a} = e^{\mu}_{a}(u,\xi) A_{\mu}, \;\; A_{i} = \frac{\sqrt{2}}{M}
n^{\mu}_{i}(u) A_{\mu}. \]
Here
\[ e^{\mu}_{a}= \frac{\partial x^{\mu}}{\partial u^a} = Y^{\mu}_{,a}
+ \xi^i K^{ib}_{a} Y^{\mu}_{,b} + \epsilon^{ij}\omega_{a}\xi^{i}
n^{\mu}_{j}. \]
Notice that the formula defining the components $A_i$ contains the factor
$\sqrt{2}/M$, so that $A_i$  are dimensionless, similarly as $s^i$
and $\phi$. We do not rescale the coordinates $u^a$ and the
components $A_a$. The variables $\xi^i$ present in $G^{\alpha \beta}$ and
$\sqrt{-G}$ are also rescaled by the same factor. We shall use the
following notation for the U(1) gauge covariant derivatives
\[ D_a = \frac{\partial}{\partial u^a} + i q A_a, \;\;
 D_i = \frac{\partial}{\partial s^i} + i q A_i.    \]

 Using the rescaled co-moving coordinates and the rescaled fields we
can write Eq.(2) in the following form
\begin{equation}
    D_{i}D_{i}\phi + \phi - |\phi|^2 \phi = -\frac{\partial_i h}{h}
D_i \phi
\end{equation}
 \[  + \frac{2}{M^2} \frac{1}{\sqrt{-g}h} (D_{a} + \epsilon^{il} s^{l}
  \omega_{a} D_i)  \left[ \sqrt{-g}hG^{ab}
 ( D_b \phi +\epsilon^{kr}s^{r}\omega_{b}D_{k}\phi)\right].  \]
The function $h$ is given by formula (8); $\; \partial_i$ denotes
$ \partial / \partial s^i$.

The Maxwell equation (3) in the co-moving coordinates has the  form
\begin{equation}
 \partial_{\alpha}(\sqrt{-G} G^{\alpha \beta} F_{\beta \gamma}) =
i q \sqrt{-G}  (\Phi^{\ast} D_{\gamma}\Phi - \Phi D_{\gamma} \Phi^{\ast}),
\end{equation}
where
\[ F_{\beta \gamma} = \frac{\partial A_{\gamma}}{\partial \zeta^{\beta}}
- \frac{\partial A_{\beta}}{\partial \zeta^{\gamma}}. \]
It is convenient to split Eq.(13) into two equations and to pass to the
rescaled variables. In the case $\gamma =k =1,2$ Eq.(13) gives
 \begin{equation}
 \lambda \partial_{i} F_{ik} + 2 i q (\phi^{\ast} D_{k}\phi -
\phi D_{k}\phi^{\ast}) = - \lambda \frac{\partial_i h}{h}F_{ik}
\end{equation}
\[  + \frac{2 \lambda}{M^2} \frac{1}{\sqrt{-g}h}
 ( \partial_{a} + \epsilon^{il} s^{l} \omega_{a} \partial_i )
    \left[ \sqrt{-g}hG^{ab}
   ( F_{bk} +  \epsilon^{sr} s^{r} \omega_{b}  F_{sk}) \right], \]
where
\[ F_{ik} = \partial_{i}A_{k} - \partial_{k}A_{i},\; \;
   F_{bk} = \partial_{b}A_{k} - \partial_{k}A_{b}. \]
For $\gamma = c = 0, 3$ we obtain from (13)
\begin{equation}
  \lambda \partial_{i} F_{ic} + 2 i q (\phi^{\ast} D_{c}\phi -
\phi D_{c}\phi^{\ast}) = - \lambda \frac{\partial_i h}{h}  F_{ic}
\end{equation}
 \[ + \frac{2 \lambda}{M^2} \frac{1}{\sqrt{-g}h}
 ( \partial_{a} +  \epsilon^{ik}s^{k}\omega_{a} \partial_i  )
 \left[ \sqrt{-g}hG^{ab}
   ( F_{bc}  +\epsilon^{tl}s^{l}\omega_{b}F_{tc})\right]. \]
Equations (12), (14), (15) have the form convenient for constructing the
perturbative expansion in powers of $1/M$.

The reason for the splitting of the Maxwell equations into the two groups
is that the components $A_{i}$ play different dynamical role in the motion
of the vortex than $A_{c}$. The components $A_i$, determined basically
from Eq.(14), do not vanish for any vortex and they are present even in the
zeroth order approximation. The components $A_c$ are determined from Eq.(15).
They vanish in the zeroth and first order approximations for
all vortices, and for some vortices to all orders (e.g. for a
straight-linear vortex). Moreover, we shall see that in the $1/M$ expansion
we first find the n-th order contributions to $\phi$ and $A_i$ and the n-th
order contribution to $A_c$ is calculated afterwards.

The co-moving coordinate system in general does not cover the whole
space-time. We assume that the range of its validity includes the physically
most interesting region -- outside of it the fields are exponentially close
to the vacuum. The range of validity of the co-moving coordinates can be
determined from the condition $h(u^a, \xi^i) >0$.  We will not discuss
here this purely mathematical point.

\section{ Expansion in the width -- the zeroth order solution and
zero modes}

The present Section is devoted to a preliminary discussion of the expansion
in the width for the vortex solution constructed with the help of
Hilbert-Chapman-Enskog method. The zeroth order solution we obtain
coincides with the one known from earlier
applications of the expansion in the width to vortices, \cite{3}.
In the next Section we calculate the first order contribution.

We seek approximate solutions of Eqs.(12),(14),(15) in the form of power
series in $1/M$:
\begin{equation}
\phi = \phi^{(0)} + \frac{1}{M} \phi^{(1)} +
\frac{1}{M^2} \phi^{(2)} +... \;,
\end{equation}
\begin{equation}
A_{i} = A^{(0)}_{i} + \frac{1}{M} A^{(1)}_{i} + \frac{1}{M^2} A^{(2)}_{i}
 +...\; ,
\end{equation}
\begin{equation}
A_{a} = A^{(0)}_{a} + \frac{1}{M} A^{(1)}_{a} + \frac{1}{M^2} A^{(2)}_{a}
 +... \;.
\end{equation}
We also assume that these fields do not have any component oscillating in
$\tau$ with a frequency proportial to $M$. This assumption ensures that
when acting with the derivative $\partial_{\tau}$ on the n-th order
contribution we obtain still the n-th order contribution. To illustrate
the point: the $\tau$-derivative of the n-th order function
$M^{-n} \cos M\tau$ belongs to the (n-1)-th order.
This assumption has a physical content -- it eliminates a whole class of
solutions, in particular the ones describing radiation of the Higgs and
vector particles. In order to include such solutions one would have to
construct an extended perturbative scheme, e.g., in analogy to
considerations presented in \cite{7}, where such oscillating component was
taken into account in the case of domain walls in the framework
of a polynomial approximation.

In the perturbative calculations we would like to preserve the local
U(1) gauge invariance of the Abelian Higgs model. To this end, we use
the trick borrowed from background field techniques in field theory: we
assume that the gauge transformations act only on $A^{(0)}_{\alpha}$, while
$A^{(n)}_{\alpha}$ with $n>0$ are gauge invariant. Then, the total vector
field $A_{\alpha}$ has the right transformation law. As for the Higgs field,
each contribution $\phi^{(n)}$ is multiplied by the same
$(\zeta^{(\alpha)})$-dependent phase factor.  Then, in the gauge covariant
derivatives it is sufficient to use the zeroth order vector field
$A^{(0)}_{\alpha}$, i.e.
\[ D^{(0)}_{a} \phi = (\frac{\partial}{\partial u^a} +
i q A^{(0)}_{a}) \phi, \;\;  D^{(0)}_{k} \phi =
 (\frac{\partial}{\partial s^k} +
i q A^{(0)}_{k}) \phi. \]

As explained in detail in paper \cite{1}, the expansion in $1/M$ is an
example of singular perturbations. In order to calculate the perturbative
contributions correctly we use the Hilbert-Chapman-Enskog method. In this
method, apart from obtaining the solution in a given order it is essential
to look also at the equations in the higher orders -- from them one obtains
consistency conditions for the solution in the given order. The consistency
condition for the zeroth order solution follows from the equations in the
first order. The equations in the zeroth and first orders are obtained by
inserting the expansions (16-18) into Eqs.(12),(14),(15) and collecting the
terms of the order $(1/M)^0$ and $(1/M)^1$.

In the zeroth order, Eqs.(12),(14),(15) are  reduced to the following
equations:
\begin{equation}
D^{(0)}_{i} (D^{(0)}_{i} \phi^{(0)})  +  \phi^{(0)}  -
|\phi^{(0)} |^2 \phi^{(0)} = 0,
\end{equation}
\begin{equation}
\lambda \partial_{i} F_{ik}^{(0)} = - 2 i q
[ \phi^{\ast(0)} (D^{(0)}_{k} \phi^{(0)}) - \phi^{(0)}
 (D^{(0)}_{k} \phi^{(0)})^{\ast}],
\end{equation}
\begin{equation}
\lambda \partial_{i} F_{ic}^{(0)} = - 2 i q
[ \phi^{\ast(0)} (D^{(0)}_{c} \phi^{(0)}) - \phi^{(0)}
 (D^{(0)}_{c} \phi^{(0)})^{\ast}].
\end{equation}
Equations (19),(20) coincide with equations for the static
Abrikosov-Nielsen-Olesen (A-N-O) vortex \cite{8,9}.

Let us quote here basic facts about the A-N-O vortex. We shall use them in
the next Section. The axially symmetric Ansatz for the A-N-O vortex has the
form
\begin{equation}
\phi^{(0)} = e^{in\theta} F(s), \; A^{(0)}_{i} = \epsilon^{ik} \frac{s^k}{s}
(\frac{n}{qs} - H(s) ),
\end{equation}
where $s=\sqrt{(s^1)^2 + (s^2)^2},\; \theta$ is the azimuthal angle in the
$(s^1,s^2)$-plane, and $n$ is the topological charge of the vortex. This
Ansatz reduces Eqs.(19),(20) to the following equations
\begin{equation}
F'' + \frac{1}{s} F' -q^2 H^2 F + F - F^3 =0,
\end{equation}
\begin{equation}
B'(s) = \frac{4q^2}{\lambda} F^2 H,
\end{equation}
where
\[ B = H' + \frac{H}{s},  \]
and ' denotes the derivative $d/ds$. These equations are supplemented by the
boundary conditions for the functions $F, H$:
\begin{equation}
 F(0) = 0,\; F(\infty) = 1,
 \end{equation}
\begin{equation}
\lim_{s \rightarrow 0} s H(s) = \frac{n}{q}, \; H(\infty) = 0.
\end{equation}
We will call  the functions $F, H$ obeying Eqs.(23), (24) with the
 boundary conditions (25), (26) the A-N-O functions, and denote them by
 $F_0 (s), H_0 (s)$. They exponentially vanish for large s.
The magnetic field of the vortex is equal to
\begin{equation}
F^{(0)}_{ik} = \epsilon^{ik} ( \frac{H_0}{s} +H_{0}') \equiv \epsilon^{ik}
B_0 (s).
\end{equation}
Explicit, exact analytical form of the A-N-O functions is not known.
Nevertheless, they are very well investigated, see, e.g. \cite{10}.

For $\phi^{(0)}, A^{(0)}_{i}$ given by the A-N-O solution, equation (21)
is reduced to
\begin{equation}
\triangle A^{(0)}_c - \frac{4q^2}{\lambda} F^2_0(s) A^{(0)}_c = 0,
\end{equation}
where $\triangle \equiv \partial_i \partial_i$ is the 2-dimensional
Laplacian. The operator on the l.h.s. of this equation does not have
non-zero regular, localised solutions. This can be easily
seen by interpreting the operator $ - \triangle + \frac{4q^2}{\lambda}
F^2_0$ as a Schroedinger operator with the potential $\frac{4q^2}{\lambda}
F^2_0$. Because this potential is non-negative, there are no bound states
with zero eigenvalues. Therefore,
\begin{equation}
A^{(0)}_{c} = 0.
\end{equation}
Notice that Eq.(28) coincides with Eq.(24) for $H(s)$. The fact that the
solution $H_{0}(s)$ of Eq.(24) is non-zero is not in contradiction with
the conclusion (29) because $H_0$ is singular at $s=0$ (see (26)), while
$A^{(0)}_{c}$  is required to be regular for all $s^i$ and exponentially
vanishing for large $s$. These regularity conditions follow from
the requirement that the fields $F_{\alpha\beta}$ should be regular and
exponentially localised.

Notice that by adopting the concrete form of the zeroth order solution
we implicitely fix the gauge for two local gauge symmetries. The
first local gauge group is formed by the $(x^\mu )$-dependent
"electromagnetic" U(1) transformations. The other one consists of
$(u^a)$-dependent SO(2) rotations in the $(\xi^1, \xi^2)$-plane, equivalent
to SO(2) rotations of the basis $(n_1, n_2)$ in the plane orthogonal to the
world-sheet $\Sigma$ of the string. The third local gauge symmetry, namely
invariance with respect to reparametrisations of the world-sheet of the
string remains explicitely preserved. In the present paper we
prefer to work with the concrete gauge fixing for the former two gauge
symmetries. Actually, one can provide a formulation of the Abelian Higgs
model in the co-moving coordinates which is explicitely invariant under
the three local gauge symmetries, \cite{11}.

Now we will turn to the consistency conditions for the zeroth order
solution. To this end we have to derive equations in the first order.
It is convenient to introduce a compact notation for the fields and the
field equations. Thus,
\begin{equation}
\Psi = \left(
\begin{array}{c}
 \phi \\
 \phi^{\ast} \\
 A_{k}
 \end{array}
 \right).
 \end{equation}
The compact notation for the left hand side of the set of equations formed
by Eq.(12), its complex conjugate counterpart, and Eq.(14) is
${\cal F}(\Psi)$. Then, the zeroth order equations (19), (20) can be
written as
\begin{equation}
{\cal F}(\Psi^{(0)}) = 0,
\end{equation}
\begin{equation}
\Psi^{(0)} = \left(
\begin{array}{c}
 \phi^{(0)} \\
 \phi^{\ast (0)} \\
 A_{k}^{(0)}
 \end{array}
 \right).
 \end{equation}
In this notation we regard $\phi$ and $\phi^{\ast}$ as two independent
fields obeying Eq.(12) and its complex conjugate countertpart,
correspondingly. This is equivalent to considering the real and imaginary
parts of the field $\phi$ and  Eq.(12). Notice that we have left
 aside Eq.(15) -- it will be
solved later. This is possible because $A_a$ is not present on the l.h.s.
of Eqs.(12),(14).

Let us now consider a small correction
 \begin{equation}
\delta\Psi = \left(
\begin{array}{c}
\delta\phi \\
\delta\phi^{\ast} \\
\delta A_{k}
 \end{array}
 \right)
 \end{equation}
to the zeroth order solution $\Psi^{(0)}$. Because of Eq.(31) we have
 \begin{equation}
 {\cal F}(\Psi^{(0)} + \delta \Psi) = \hat{L} \delta \Psi +
 {\cal O}((\delta \Psi)).
\end{equation}
Here $\hat{L}$ is a linear operator obtained by expanding the l.h.s. of
Eqs.(12),(14) around $\phi^{(0)}, A^{(0)}_{i}$. Including also $\phi^{\ast}$
and the complex conjugate of Eq.(12) we find that

\begin{eqnarray}
\hat{L} =
   \left[
\begin{array}{c}
D^{(0)}_{i} D^{(0)}_{i} + 1 - 2 |\phi^{(0)}|^2     \\
-(\phi^{\ast (0)})^2      \\
2 i q [ \phi^{\ast (0)} D^{(0)}_{k}  - ( D^{(0)}_{k}\phi^{(0)} )^{\ast} ]
\end{array}    \right.   &
 \begin{array}{c}
   -(\phi^{(0)})^2   \\
 D^{\ast (0)}_{i} D^{\ast (0)}_{i} + 1 -  2 |\phi^{(0)}|^2    \\
 - 2 i q [ \phi^{(0)} D^{\ast (0)}_{k}  - ( D^{(0)}_{k}\phi^{(0)}) ]
\end{array} \nonumber  \\
    &    \nonumber    \\
 &   \left.
\begin{array}{c}
2 i q ( D^{(0)}_{l}\phi^{(0)} ) + i q \phi^{(0)} \partial_{l}  \\
   - 2 i q ( D^{(0)}_{l}\phi^{(0)} )^{\ast}
 - i q \phi^{\ast (0)} \partial_{l}  \\
(\lambda \partial_i \partial_i - 4 q^2  |\phi^{(0)}|^2)
 \delta^{kl}  - \lambda \partial_k \partial_l
\end{array} \right].
\end{eqnarray}

The equation for the first order correction to $\Psi^{(0)}$, i.e. for
\begin{equation}
\Psi^{(1)} = \left(
\begin{array}{c}
 \phi^{(1)} \\
 \phi^{\ast (1)} \\
 A_{k}^{(1)}
 \end{array}
 \right),
 \end{equation}
easily follows from formula (34) and Eqs.(12),(14). It has the following
form
 \begin{equation}
 \hat{L}\Psi^{(1)} = - \sqrt{2}  \left(
 \begin{array}{c}
 D^{(0)}_{i}\phi^{(0)}  \\
(D^{(0)}_{i}\phi^{(0)})^{\ast}  \\
 \lambda   F^{(0)}_{ik}
\end{array}
\right) K^{ia}_{a},
\end{equation}
where the r.h.s. has been obtained by expanding the r.h.s. of Eqs.(12), (14).
Equation (37) will be solved in the next Section.

{}From Eq.(37) we obtain the consistency conditions for the zeroth order
solution. The crucial observation is that there are left eigenvectors of
 operator $\hat{L}$ with the corresponding eigenvalue equal to zero,
 i.e. normalizable, non-zero $\Psi^{l.z.m.}$ such that
\begin{equation}
\int ds^1 ds^2 (\Psi^{l.z.m.})^{\ast T} \hat{L} \Psi = 0
\end{equation}
for all $\Psi$ vanishing sufficiently quickly for large $s$. In condition
(38) $T$ denotes the matrix transposition, and the abbreviation l.z.m.
stands for "left zero mode". It is easy to check that the operator $\hat{L}$
is Hermitean with respect to the scalar product given by
\begin{equation}
< \Psi_1 | \Psi_2 > = \int ds^1 ds^2 \Psi^{\ast T} \Psi,
\end{equation}
in the space of $\Psi$'s which have the second component equal to the
complex conjugation of the first one. (Therefore, it is a Hilbert space
over real numbers and not over complex numbers.) Actually, the Higgs
self-coupling constant $\lambda$ has been placed in front of the
$F_{\alpha \beta}$ field in Eqs.(14),(15) just in order to obtain
$\hat{L}$ Hermitean with respect to the simple scalar product (39). In
the case of Hermitean operator the l.z.m.'s coincide with right zero
modes, which are defined as normalizable (non-zero) solutions of
the equation
\begin{equation}
\hat{L} \Psi^{r.z.m.} = 0.
\end{equation}

The zero modes can be easily found with the help of formula (34) by using a
particular $\delta \Psi$ corresponding to symmetries of Eqs.(19),(20).
These equations are regarded here as equations for functions on
the plane $(s^1, s^2)$. Because of invariance of Eqs.(19),(20) with respect
to translations in the $(s^1, s^2)$-plane, one can generate from the
solution $\Psi^{(0)}$ infinitely many other solutions by applying the
translations. We shall denote these new solutions by $T_{a}\Psi^{(0)}$.
Thus,
\[   T_{a}\Psi^{(0)} = \left(
\begin{array}{c}
\phi^{(0)}(s^i + a^i)  \\
\phi^{\ast (0)}(s^i + a^i)  \\
 A^{(0)}_{k}(s^i + a^i)
 \end{array}
 \right), \]
where $a = (a^i)$ is a constant vector defining the translation. It is
clear that
 \[     {\cal F}( T_{a}\Psi^{(0)} ) = 0.  \]
Taking infinitesimal $a$, we obtain from formula (34) that
\[ \hat{L} \Psi_{l} = 0, \]
where
\[ \Psi_{l} = \frac{\partial ( T_{a}\Psi^{(0)} )}{\partial a^l}
 \left|_{a=0}  \right. =  \left(   \begin{array}{c}
 \frac{\partial \phi^{(0)}}{\partial s^l}  \\
 \frac{\partial \phi^{\ast (0)}}{\partial s^l}  \\
 \frac{\partial A_{k}^{(0)}}{\partial s^l}
 \end{array}   \right).
 \]
These $\Psi_{l}$ can not be used as the zero modes because they are not
normalizable. The correct zero modes are obtained by combining the
translation with a gauge transformation in the spirit of improved
variations of Jackiw and Manton, \cite{12}. Instead of $T_{a}\Psi^{(0)}$
we take
\begin{equation}
 \tilde{T}_{a}\Psi^{(0)} = \left(
\begin{array}{c}
    e^{i \chi(a; s^k)} \phi^{(0)}(s^i + a^i)  \\
   e^{-i \chi(a; s^k)}   \phi^{\ast (0)}(s^i + a^i)  \\
 A^{(0)}_{k}(s^i + a^i) - \frac{1}{q} \partial_k \chi(a;s^i)
 \end{array}
 \right),
 \end{equation}
where $\chi(a;s^i) = q a^l A_l^{(0)}(s^i)$. It is clear that also
$ \tilde{T}_{a}\Psi^{(0)}$ is a solution of Eqs.(19),(20). As the zero
modes we take
\begin{equation}
 \Theta_{l} = \frac{\partial (\tilde{T}_{a}\Psi^{(0)} )}{\partial a^l}
 \left|_{a=0}  \right. =  \left(   \begin{array}{c}
 D^{(0)}_l \phi^{(0)}  \\
 (D^{(0)}_l \phi^{(0)})^{\ast}  \\
 F^{(0)}_{lk}
 \end{array}   \right).
 \end{equation}
 where $l=1,2.$
 They are normalizable and orthogonal -- one can easily check that
\[ < \Theta_{k} | \Theta_{l} > = \delta^{kl} \int ds^1 ds^2 ( B_{0}^{2}
 + F_{0}^{'2}  + q^2 H_{0}^{2}F^{2}_{0}  ),  \]
 where $F_0,H_0,$ and $B_0$ are the A-N-O functions.

There also exists  a solution of Eq.(40) related to the U(1) gauge
invariance. It has the form
\[ \Psi_{g} = \left(
 \begin{array}{c}
i \phi^{(0)}  \\
 - i \phi^{\ast(0)}  \\
 0
 \end{array}   \right).  \]
Another possible solution, corresponding to rotations around the
origin in the $(s^1,s^2)$-plane is in fact identical with $\Psi_{g}$
because in the case of the A-N-O vortex the rotations are equivalent to
global U(1) gauge transformations. Notice however that   $\Psi_{g}$ is
not normalizable with respect to the scalar product (39). Therefore, it
can not be accepted as the zero mode.

The consistency conditions follow from Eq.(37) by integrating it with the
zero modes $ \Theta_{l}$, like in formula (38). The r.h.s.
of Eq.(37) in general is not a linear combination of the zero modes because
of the factor $\lambda$ in front of $F_{lk}^{(0)}$. Nevertheless, the
integrations are simple and  we obtain the following conditions
\begin{equation}
K^{la}_{a} = 0,
\end{equation}
where $l=1,2$ and $a=0,3.$

The conditions (43) are equivalent to Nambu-Goto equation for a
rela\-tiv\-istic string whose world-sheet is given by the
surface $\Sigma$ in Minkowski space-time.

To summarize, the zeroth order solution is given by formulae (22) with
$F(s), H(s)$ equal to the A-N-O functions $F_{0}(s), H_{0}(s)$. The
components $A^{(0)}_{a}$ vanish. The variables $s = M \xi^i$ are related to
the Cartesian coordinates by formula (5). For the functions
 $Y^{\mu}(u^a)$ in that formula we can take any solution of Nambu-Goto
equation (43). In the zeroth order approximation the string coincides
with the zeros of the Higgs field.

\section{The first order corrections}

The first order corrections to $\phi$ and $A_k$ obey Eq.(37) with vanishing
the r.h.s., due to the consistency conditions (43). This equation is
satisfied by linear combinations of the zero modes and $\Psi_{g}$
with   coefficients which do not dependent on $s^i$  , i.e.
\begin{equation}
\Psi^{(1)} = C^{l}(u^a) \Theta_{l}(s^i) + C^{3}(u^a) \Psi_{g}(s^i),
\end{equation}
where $l=1,2$. Now we have included $\Psi_{g}$ because in Eq.(37) there are
no integrals. The functions $C^l, C^3$ are real, otherwise the second
component of of $\Psi^{(1)}$ would not be equal to the complex conjugation
of the first one.  In the following we shall show that the functions
 $C^l, C^3$  are not arbitrary.

We also have to solve the first order equation obtained from Eq.(15) for
$A_b$. With the Nambu-Goto equations (43) and the zeroth order results
taken into account, it has the following form
\begin{equation}
\triangle A^{(1)}_b - \frac{4q^2}{\lambda} F^2_0(s) A^{(1)}_b = 0.
\end{equation}
We already know from the discussion following Eq.(28) that the only regular,
localised solution of Eq.(45) is the trivial one,
\begin{equation}
A^{(1)}_{b} = 0.
\end{equation}

By the analogy with the case of domain walls \cite{1} we expect consistency
conditions for the first order solution coming from the second and third
order equations. These conditions have
the form of equations which must be obeyed by the functions $C^l, C^3$.
Therefore, we have to consider the perturbative equations in the second and
third orders. As for still higher orders, we expect that the corresponding
consistency conditions can be "saturated" by new arbitrary functions,
introduced in each order as the general solution of the homogeneous equation
$\hat{L}\Psi^{(n)} =0$. All such solutions have the form (44) with the
functions $C^l, C^3$ replaced by a new set of functions in each order.
They have to be added to a solution of the full perturbative equation in the
given order to obtain the general solution in the n-th order.  The
full perturbative equations in the second and third orders are given below,
see Eqs.(47),(52).

The equation in  the second order has the following form
\begin{equation}
 \hat{L}\Psi^{(2)} =   \left(
 \begin{array}{c}
f^{(2)}  \\
f^{\ast (2)}  \\
 a^{(2)}_{k}
\end{array}
\right),
\end{equation}
where
\begin{eqnarray}
& f^{(2)} = 2 \phi^{(0)} | \phi^{(1)}|^2 +  (\phi^{(1)})^2 \phi^{\ast (0)} +
q^2 A_{i}^{(1)} A_{i}^{(1)} \phi^{(0)}  &   \nonumber    \\
 &  - 2 i q  A_{i}^{(1)} D^{(0)}_{i}\phi^{(1)} -
 i q (\partial_{i}  A_{i}^{(1)})  \phi^{(1)} + 2 K^{ib}_{a} K^{ja}_b s^j
 D^{(0)}_i \phi^{(0)}   &    \\
 & + \frac{2}{\sqrt{-g}} (\partial_a + \epsilon^{il} s^l \omega_a D^{(0)}_i)
 ( \sqrt{-g} g^{ab} \epsilon^{kr} s^r \omega_b  D^{(0)}_k \phi^{(0)} ), &
 \nonumber
\end{eqnarray}
and
\begin{eqnarray}
& a^{(2)}_k = - 2 i q (\phi^{\ast (1)} D^{(0)}_k \phi^{(1)} -
\phi^{(1)} D^{(0)}_k \phi^{\ast (1)} )    &   \nonumber    \\
 & + 2 q^2 A^{(1)}_k ( \phi^{\ast (1)}  \phi^{(0)}
+ \phi^{(1)} \phi^{\ast (0)} ) +
 2 \lambda K^{ib}_{a} K^{ja}_b s^j F^{(0)}_{ik} &  \\
 & + \frac{2\lambda}{\sqrt{-g}} (\partial_a +
  \epsilon^{il} s^l \omega_a \partial_i)
 ( \sqrt{-g} g^{ab} \epsilon^{tr} s^r \omega_b F_{tk}^{(0)} ). &
 \nonumber
\end{eqnarray}
We also have the following equation for $A^{(2)}_c$ (obtained by expanding
Eq.(15))
\begin{eqnarray}
 & \triangle A^{(2)}_c - \frac{4q^2}{\lambda} F^2_0(s) A^{(2)}_c =
 \partial_c (\partial_i A^{(2)}_i)  &  \\
  & - \frac{ 2 i q}{\lambda} (\phi^{\ast (0)} \partial_c \phi^{(2)} -
\phi^{(0)} \partial_c \phi^{\ast (2)} +
 \phi^{\ast (1)} \partial_c \phi^{(1)} -
 \phi^{(1)} \partial_c \phi^{\ast (1)} ). & \nonumber
\end{eqnarray}
In the formulae for $f^{(2)}, a^{(2)}_k$, and on the r.h.s. of Eq.(50) the
zeroth order results and the Nambu-Goto equation (43) have already been
taken into account.

{}From  Eq.(47) one can in principle determine the second order correction
$\Psi^{(2)}$. In practice we do not hope to obtain an explicit formulae
for $\Psi^{(2)}$ -- we do not even know the explicit form of the zeroth order
functions $\phi^{(0)}, A_k^{(0)}$ which are present in the
operator $\hat{L}$. On the other hand, the functions $\phi^{(0)}, A_k^{(0)}$
are known numerically, and the coefficients and the r.h.s. in Eq.(47) are
regular. Therefore, we expect that it will be possible to investigate the
solutions of this equation by numerical methods.  This task we will leave
for a separate work. In the present paper we would like to focus on the
description of the perturbative scheme and on the first
order correction  $\Psi^{(1)}$.

As explained in Section 3, the consistency conditions are obtained by
integrating the both sides of the perturbative equation, i.e. Eq.(47) in
the present case, multiplied by the zero modes $\Theta_{l}$:
\[ 0 = \int ds^1 ds^2 [ (D^{(0)}_{l}\phi^{(0)})^{\ast} f^{(2)} +  c.c.
+ F^{(0)}_{lk} a^{(2)}_{k} ]   \]
(c.c. stands for "complex conjugate").  Most contributions
 to the integral vanish because the integrands are odd functions of
 the variables $s^i$ and the integration range is from  $-\infty$ to
 $\infty$. The non-vanishing terms on the r.h.s. lead to the following
 condition
 \[ 0 = - d_0 q \;\epsilon^{lk} C^{l}(u^a) C^{3}(u^a),   \]
where $l,k=1,2$, and $d_0$ is a non-vanishing constant given in the
Appendix B.

We shall see that in general $C^l$ are not equal to zero. Therefore, we put
\begin{equation}
C^3 = 0.
\end{equation}

Equation (50) does not yield any restrictions on the functions $C^l, C^3$
because the operator on the l.h.s. of it (the same as in Eq.(28)) does not
have zero eigenvalues.

Further consistency conditions follow from the equations in the third order.
These conditions have the form of non-linear, intercoupled, inhomogeneous
wave equations for $C^{l}(u^a)$ regarded as fields on the world-sheet
$\Sigma$ of the Nambu-Goto string. Unfortunately, now the calculations become
rather cumbersome -- one could hardly expect simple formulae in the third
order approximation in the model with six coupled fields. The equation
for the third order correction $\Psi^{(3)}$ has the form
  \begin{equation}
 \hat{L}\Psi^{(3)} =   \left(
 \begin{array}{c}
f^{(3)}  \\
f^{\ast (3)}  \\
 a^{(3)}_{k}
\end{array}
\right),
\end{equation}
where
\[ f^{(3)} = f_1^{(3)} + f_2^{(3)},\;\; a^{(3)}_{k} =
a^{(3)}_{1k} + a^{(3)}_{2k}.  \]
Here $f_1^{(3)},\; a^{(3)}_{1k}$ ( $f_2^{(3)},\; a^{(3)}_{2k}$ ) denote
the terms obtained by expanding the l.h.s.'s ( the r.h.s.'s ) of Eqs.(12),
(14), respectively. We find that

\begin{eqnarray}
& f_1^{(3)} = 2 \phi^{(0)} \phi^{(1)}  \phi^{\ast (2)} +
2 ( \phi^{\ast (1)} \phi^{(0)} + \phi^{\ast (0)} \phi^{(1)} ) \phi^{(2)}
 +  \phi^{ (1)} | \phi^{(1)}|^2  &   \nonumber   \\
 &  - 2 i q  A_{i}^{(1)} D^{(0)}_{i}\phi^{(2)}
- 2 i q  A_{i}^{(2)} ( D^{(0)}_{i}\phi^{(1)} +
i q  A_{i}^{(1)} \phi^{(0)} )    & \\
&- i q (\partial_{i} A_{i}^{(1)}) \phi^{(2)} - i q (\partial_{i} A_{i}^{(2)})
   \phi^{(1)},  &  \nonumber
\end{eqnarray}

\begin{eqnarray}
& a^{(3)}_{1k} =  - 2 i q [\phi^{\ast (2)} D^{(0)}_k \phi^{(1)} -
\phi^{(2)} ( D^{(0)}_k \phi^{(1)} )^{\ast} & \nonumber \\
&  +  \phi^{\ast (1)} D^{(0)}_k \phi^{(2)} -
\phi^{(1)} ( D^{(0)}_k \phi^{(2)} )^{\ast} ]    &       \\
 & + 4 q^2 A^{(1)}_k ( \phi^{\ast (2)}  \phi^{(0)} +
 \phi^{(2)}  \phi^{\ast (0)} )  + 4 q^2 A^{(2)}_k
 ( \phi^{\ast (1)}  \phi^{(0)} +   \phi^{(1)}  \phi^{\ast (0)} ). &
 \nonumber
 \end{eqnarray}
$\;\;\;$ \\
and

\begin{eqnarray}
&  f^{(3)}_2 =   + 2 K^{ib}_{a} K^{ja}_b s^j ( D^{(0)}_i \phi^{(1)} + i q
  A^{(1)}_i \phi^{(0)} )   & \nonumber \\
  &   + 2 i q \epsilon^{it} s^t \epsilon^{kr} s^r
  \omega_a  \omega_b g^{ab}(D^{(0)}_k \phi^{(0)}) A^{(1)}_i &       \\
&  + \frac{2}{\sqrt{-g}} (\partial_a +
  \epsilon^{it} s^t \omega_a D^{(0)}_i) [ \sqrt{-g} g^{ab}
  ( \partial_b \phi^{(1)} +  \epsilon^{kr} s^r \omega_b
  ( D^{(0)}_k \phi^{(1)} + i q A^{(1)}_k \phi^{(0)} ) ) ]
  & \nonumber   \\
& - \frac{4 \sqrt{2}}{\sqrt{-g}} (\partial_a +
 \epsilon^{it} s^t \omega_a D^{(0)}_i)
( \sqrt{-g} K^{pab} \omega_b s^p  \epsilon^{kr} s^r D^{(0)}_k \phi^{(0)} ),
 & \nonumber
 \end{eqnarray}

\begin{eqnarray}
 & a^{(3)}_{2k}  = 2 \lambda K^{ib}_{a} K^{ja}_b s^j F^{(1)}_{ik}
  &  \nonumber \\
& + \frac{2\lambda}{\sqrt{-g}} (\partial_a +
 \epsilon^{il} s^l \omega_a \partial_i) [ \sqrt{-g} g^{ab}
( F^{(1)}_{bk} + \epsilon^{tr} s^r \omega_b F_{tk}^{(1)} ) &  \\
& - 2 \sqrt{2} \sqrt{-g} K^{pab} \omega_b s^p
\epsilon^{tr} s^r F^{(0)}_{tk} ]. &
 \nonumber
\end{eqnarray}
In these formulae we have taken into account the Nambu-Goto equation as well
as the fact that $A^{(1)}_b$ vanishes.

In the third order we also have an equation for $A^{(3)}_c$. We will not
present it here because it does not lead to consistency conditions -- the
operator on the l.h.s. of it is the same as in Eqs.(28),(45).

The next step is to integrate  over $s^1, s^2$ the r.h.s. of Eq.(52)
multiplied by the zero modes. At first look this might seem a difficult
task because the second order corrections $\phi^{(2)}, A^{(2)}_k$ are
given only implicitely by Eq.(47). The other functions, including the
zero modes, are explicitely expressed by the A-N-O functions $F_0(s),
H_0(s)$. Luckily, it turns out that the $1/M$ expansion has the special
property that the troublesome integrals can be expressed by integrals
involving only the zeroth and first order corrections, which are explicitely
given in terms of the A-N-O functions. Derivation of this relation is
based on the translational and U(1) gauge invariances of the model, and it
strongly reminds derivations of Ward identities for Green functions in
quantum field theory. In the quantum case, instead of the integration over
$s^1, s^2$ there is a functional integral over fields. For this reason,
we will call the identity presented below the classical Ward identity.

All the terms with $\phi^{(2)}, \phi^{\ast (2)}, A^{(2)}_k $ present on
the r.h.s. of Eq.(52) are obtained by expanding the l.h.s.'s of
Eqs.(12),(14). They are collected as $f^{(3)}_1$ and $a^{(3)}_{1k}$.
The crucial observation is that they follow from  the expression
\[- \hat{L} ( \Psi^{(0)} + \frac{1}{M} \Psi^{(1)} ) \Psi^{(2)} \]
by expanding it with respect to $1/M$ and keeping the term linear in $1/M$
 (with $1/M$ dropped out).  The operator
 $\hat{L} ( \Psi^{(0)} + \frac{1}{M} \Psi^{(1)} ) $ is given by formula (35)
 with $\Psi^{(0)}$ replaced by $ \Psi^{(0)} + \frac{1}{M} \Psi^{(1)}$. The
 integrals we are discussing have the form
 \[ {\cal I}_l \equiv -  \int ds^1 ds^2
  \Theta_{l}^{\dagger}\; \left.  \frac{\partial \hat{L} ( \Psi^{(0)}
   + \Psi^{(1)}/M)}{\partial (1/M) } \right|_{1/M =0}  \Psi^{(2)},  \]
 where $l=1,2$ and $\Psi^{(1)}$ is given by formula (44) with $C^3 =0$.
 In the Appendix  A we prove that
 \begin{equation}
 {\cal I}_l = \int ds^1 ds^2 \; C^r \left(
  \begin{array}{c}
 D^{(0)}_r  D^{(0)}_l \phi^{(0)}  \\
 (D^{(0)}_r D^{(0)}_l \phi^{(0)})^{\ast}  \\
 \frac{\partial}{\partial s^r} F^{(0)}_{lk}
 \end{array}
  \right)^{\dagger}   \hat{L} ( \Psi^{(0)}) \Psi^{(2)}.
  \end{equation}
The expression $ \hat{L} ( \Psi^{(0)}) \Psi^{(2)}$ on the r.h.s. of the
classical Ward identity (57) is eliminated with the help of Eq.(47).
In this manner ${\cal I}_l$ is given by integrals involving the zeroth order
 solution and the first order corrections only.

Further steps in obtaining the consistency conditions are nothing more than
laborious calculating of the integrals present in the initial form of
these conditions which is
\[ 0 = {\cal I}_l + \int ds^1 ds^2 [ (D^{(0)}_{l}\phi^{(0)})^{\ast} f_2^{(3)}
       +  c.c. + F^{(0)}_{lk} a^{(3)}_{2k} ].   \]
The result we have obtained has the form of  the following equation for
$C^l$
\[ \Box^{(3)} C^l - 2 \omega_a g^{ab} \epsilon^{lk} \partial_b C^k  -
 \omega_{a} \omega_{b} g^{ab} C^l - \frac{1}{\sqrt{-g}} \partial_a
 (\sqrt{-g}g^{ab} \omega_b) \epsilon^{lk} C^k   \]
\begin{equation}
 -  K^{lb}_a K^{ja}_b C^j  + \frac{d_2}{d_1}  \omega_a g^{ab} (
 \epsilon^{lk} \partial_b C^k  +  \omega_{b} C^l )
 -  \frac{d_3}{d_1} (C^k C^k) C^l  =
\end{equation}
\[ \frac{d_4}{d_1} [  \epsilon^{lk} \frac{1}{\sqrt{-g}} \partial_a
(\sqrt{-g}\omega_b K^{kab}  + \omega_a \omega_b K^{lab} ]   +
 \frac{d_5}{d_1} \omega_a \omega_b K^{lab}  , \]
where
 \[  \Box^{(3)} C^l = \frac{1}{\sqrt{-g}} \partial_a
 (\sqrt{-g}g^{ab} \partial_b C^l).  \]
The  constants $d_i$, i=1,...5, are listed in Appendix B. The fact that
the coefficients in Eq.(58) are correlated is due to the implicit local
SO(2) invariance mentioned below formula (29).

Let us summarize our results. The obtained  to the first order in $1/M$
vortex solution has the following form ( in the co-moving coordinates )
\begin{equation}
\phi = e^{in\theta} [ F_0(s) + \frac{1}{M} C^i(u)
( \frac{s^i}{s} F^{'}_0(s)
 - iq \epsilon^{ik} \frac{s^k}{s} H_0(s) F_0(s) )],
 \end{equation}
\begin{equation}
A_{i} = \epsilon^{ik} \frac{s^k}{s} ( \frac{n}{q s} - H_0(s) )
   - \frac{1}{M} C^k(u) \epsilon^{ik}  B_0(s),
 \end{equation}
\begin{equation}
A_b = 0,
\end{equation}
where $C^k(u)$ obey Eq.(58) and $B_0=H^{'}_{0} +H_0/s$. The problem how to
fix initial data for equation (58) is discussed below.

{}From formulae (59),(60) one can find approximate position of the zeros
of the Higgs and vector fields. Using the fact that for small $s^i$ \cite{10}
\[  \phi^{(0)} \cong c_0 ( s^1 + i s^2 )^{|n|},  \]
where $c_0$ is a constant, we find that the n-th
order zero of $\phi^{(0)}$ at $s^1=s^2=0$ is split by the first order
correction into (n-1)-th order zero at  $s^1=s^2=0$ and a first order
zero at
\begin{equation}
s^i_{H} \approx - \frac{n}{M} C^i.
\end{equation}
For $|n| = 1$ only the zero (62) is present. Thus, the (n-1)-th
zero lies on the string and therefore follows a Nambu-Goto type trajectory.
The other zero  is shifted from the Nambu-Goto trajectory if the $C^l$
functions do not vanish.

As for the zeros of the $A_k$ field (k=1,2) , we find that the both
components vanish at the point
\begin{equation}
s^i_{A} \approx -\frac{2}{M} C^i,
\end{equation}
which is in general different from the origin and $s^i_H$.  This first order
zero is not so important because its
position can be shifted by the local U(1) gauge transformations dependent
on $s^1, s^2$.

The final problem we have to deal with is to fix the initial data for
Eq.(58) from which we  calculate the functions $C^l$, and also for the
Nambu-Goto equation (43) from which we find the functions $Y^{\mu}(u^a)$.
The input is the initial data for the vortex fields in the laboratory
frame coordinates. The string is only an auxilliary mathematical
object. Therefore, we can choose its initial position and velocity in a
convenient manner.

  For $|n|=1$ the most natural choice is that at the
initial $\tau = \tau_0$ the position and velocity of the string coincide with
the ones of the line of zeros of the Higgs field. This implies that
 at $\tau = \tau_0 $
\begin{equation}
  C^l =0, \;\; \partial_{\tau}C^l =0.
 \end{equation}
The values of the $C^l$ functions at later time depend on the r.h.s. of
Eq.(58). Only if it vanishes for all $\tau$ then $C^l$ remain equal to
zero for all $\tau$. Notice that for non-vanishing  of r.h.s. of Eq.(58)
it is necessary that the torsion coefficients $\omega_a$ do not vanish
identically.

 For $|n|>1$ the initial data can be fixed again by looking at the
  zeros of the initial Higgs field. In general, it is possible
 that the zeros are split already at the initial time and that they have
 nonvanishing relative velocity. For $|n| \geq 2$ the split zeros are of
 different order -- the initial position and velocity of the string are
 determined from the (n-1)-th order zero, while the initial values of the
functions $C^l,  \partial_{\tau}C^l$ are fixed by the initial position and
velocity of the  line on which lie the first order zeros. For $|n|=2$ we
can take any  of the two  zeros to determine the initial data for the
Nambu-Goto string.  Then, the other one will give the initial data for
$C^l$. The two choices are equivalent because Eq.(58) is symmetric with
respect to the change of sign of the functions $C^l$.

In all cases the functions $C^l(u^a)$ give deviations of the trajectory of
the first order zero of the Higgs field from the Nambu-Goto trajectory.

\section{Remarks}

In the present paper we have adapted the Hilbert-Chapman-Enskog method
to the problem of evolution of the vortex. The method, previously applied
to domain walls \cite{1}, seems to work quite well also in the case of
vortices, eventhough in the present case the calculations are more
complicated.

There are important differences between the aproach proposed in our paper
and the earlier applications of the expansion in the width to dynamics of
vortices in Minkowski space-time, \cite{3,4,5,6}. Apart from using the
Hilbert-Chapman-Enskog method in which the consistency conditions
play the prominent role, also new is the fact that the co-moving coordinate
system does not have to be tied to the zeros of the Higgs field.

Already the first order correction $\Psi^{(1)}$ reveals interesting
phenomena in the vortex dynamics, like the splitting of the zeros of
the Higgs field, or the possibility that for the $|n|=1$ vortex the zeros
do not follow the Nambu-Goto type trajectory. It would be also interesting
to know the second order correction, but it can be calculated probably
only by means of numerical methods.

Detailed discussion of properties of the obtained vortex solution, as well
as concrete examples of evolution of the vortex obtained by solving
Eqs.(43), (58) we will present in a separate paper, because the present
one already seems to be rather voluminous.

Let us end the present paper with a general remark that in our approach
to the vortex dynamics there is a quite interesting interplay of the
4-dimensional and 2-dimensional field theoretical models. First,
the zeroth order solution as well as the
zero modes are determined from Eqs.(19),(20) which can be regarded as
equations of the 2-dimensional Euclidean Abelian Higgs model. Second,
the functions $C^{l}(u)$ can be regarded as fields on the 2-dimensional
Nambu-Goto manifold $\Sigma$. This non-trivial background manifold
is to be determined from the Nambu-Goto equation (43). We expect
that in higher orders new sets of functions on  $\Sigma$ will also
 appear. In this manner, the original description of the vortex by
the 4-dimensional fields $\Phi, A_{\mu}$ would be transformed into another
description in terms of the Nambu-Goto string and the fields on its
world-sheet.

\vspace*{1cm}

 $\;\;$ \\
{\bf APPENDIX A. The derivation of the classical Ward identity (57) }
\vspace*{0.3cm}

We start from the identity
\[ 0 = \int ds^1ds^2\; (\tilde{T_a} \Theta_{k})^{\dagger}\;\hat{L}
   (\tilde{T_a} \Psi^{(0)})\; \tilde{T_a} \Psi^{(2)}, \;\;\;
\;\;\;\;\;\;\;\;\;\; (A1)    \]
valid for any $a$  independent of $s^1,s^2$. The operator
$\tilde{T_a}$ is given by formula (41).
The identity (A1) is just Eq.(38) written in terms of the translated
and gauge transformed fields. It is valid for any $\Psi^{(2)}$ with the
proviso that the integral is convergent, but here $\Psi^{(2)}$ is
of course the second order correction. We take the particular $a$ equal to
\[ a^l = \frac{C^{l}(u)}{M},  \]
where $C^l$ are the functions appearing in the first order solution (44).
Let us recall that $C^3 =0$. Next, we expand in $1/M$:
\[ \tilde{T_a} \Psi^{(0)} \cong \Psi^{(0)} + \frac{1}{M}
\Psi^{(1)}, \]
\[ \tilde{T_a} \Theta_{k} \cong  \Theta_{k} + \frac{1}{M} C^l
\left[ \begin{array}{ccc}
D^{(0)}_l  & 0 & 0 \\
0 &  D^{\ast(0)}_l  &0 \\
0 & 0 &  \frac{\partial}{\partial s^l}
\end{array}
\right]  \Theta_{k},  \]
and
\[  \tilde{T_a} \Psi^{(2)} \cong \Psi^{(2)} + \frac{1}{M}
\delta \Psi^{(2)} \]
(the precise form of $\delta \Psi^{(2)}$ is not needed). Substituting
these formulae on the r.h.s. of identity $(A1)$ we obtain up the
first order in $1/M$
\begin{eqnarray*}
\lefteqn{ 0=  \int ds^1ds^2 \; \left[ ( \Theta_{k})^{\dagger}\;\hat{L}
 (\Psi^{(0)})\;  \Psi^{(2)}  + \frac{1}{M} ( \Theta_{k})^{\dagger}\;\hat{L}
   (\Psi^{(0)}) \; \delta \Psi^{(2)} \right] }   \\
&   & +  \frac{1}{M} \int ds^1 ds^2 \; C^l
\left( \left[ \begin{array}{ccc}
D^{(0)}_l  & 0 & 0 \\
0 &  D^{\ast(0)}_l  &0 \\
0 & 0 &  \frac{\partial}{\partial s^l}
\end{array}
\right]  \Theta_{k} \right)^{\dagger} \;\hat{L}(\Psi^{(0)})\;
 \Psi^{(2)}      \\
\lefteqn{    - \frac{1}{M} {\cal I}_k       + {\cal O}(\frac{1}{M^2}). }
\end{eqnarray*}
The first integral on the r.h.s. vanishes because $\Theta_{k}$ is the
zero mode, and we obtain the identity (57).

\vspace*{0.5cm}
 $\;\;$\\
{\bf APPENDIX B. The  constants appearing in Section 4 }
\vspace*{0.3cm}

In order to obtain the formulae listed below we have used Eqs.(23),(24),
the boundary conditions (25),(26) and we have applied multiple
integration by parts.

\[ d_0 = 2 \pi \int^{\infty}_0 ds  s B_0 ( F_0^2 - F_0^4 +
F_0^{'2} + q^2 H^2_0 F_0^2 ),   \]

\[ d_1 =  \int^{\infty}_0 ds  s  ( \lambda B_0^2 + F^{'2}_0 + q^2
H_0^2 F_0^2 ), \]

\[ d_2 =    \frac{9}{2} \lambda \int^{\infty}_0 ds s B_0^2, \]

\begin{eqnarray*} & d_3 =  \frac{1}{4}  \int^{\infty}_0 ds
[ \frac{3}{2}  F^{'4}_0 + ( \frac{3}{2} - \frac{4q^2}{\lambda})
 ( q H_0 F_0 )^4
 + ( 1 - \frac{4q^2}{\lambda}) ( q F^{'}_0 H_0 F_0 )^2  &  \\
& + ( 2 q B_0 )^2 ( F^{'2}_0 + q^2 H_0^2 F_0^2 ) + \frac{2 q^4}{\lambda}
  H_0^2 F_0^4 ( 1 - F_0^2 ) +    \frac{2 q^4}{\lambda}
  H_0 F_0^3 F^{'}_0 ( \frac{4}{s} H_0 + B_0 )  ],    &
 \end{eqnarray*}

\[ d_4 =   \sqrt{2} \int^{\infty}_0 ds s^3 ( \lambda B_0^2 + 2 q^2 H_0^2
  F_0^2 ),    \]

\[ d_5 =  d_4 - 3 \sqrt{2} \lambda \int^{\infty}_0 ds s^3  B_0^2 .   \]

\end{document}